\documentclass[10pt,conference]{IEEEtran}
\IEEEoverridecommandlockouts
\usepackage{cite}
\usepackage{amsmath,amssymb,amsfonts}
\usepackage{algorithmic}
\usepackage{graphicx}
\usepackage{textcomp}
\usepackage{xcolor}
\def\BibTeX{{\rm B\kern-.05em{\sc i\kern-.025em b}\kern-.08em
    T\kern-.1667em\lower.7ex\hbox{E}\kern-.125emX}}
    
\usepackage{listings}
\usepackage{color}
\usepackage{listings, xcolor}
\usepackage{balance}
\usepackage[normalem]{ulem}

\definecolor{dkgreen}{rgb}{0,0.6,0}
\definecolor{gray}{rgb}{0.5,0.5,0.5}
\definecolor{mauve}{rgb}{0.58,0,0.82}

\usepackage{tikz}
\usepackage{booktabs}
\tikzstyle{mybox} = [draw=black, very thick, rectangle, rounded corners, inner ysep=5pt, inner xsep=5pt]

\begin{document}

\title{What Are We Really Testing in Mutation Testing for Machine Learning? A Critical Reflection}

\author{\IEEEauthorblockN{Annibale Panichella}
\IEEEauthorblockA{
a.panichella@tudelft.nl\\
\textit{Delft University of Technology} \\
Delft, The Netherlands \\
}
\and
\IEEEauthorblockN{Cynthia C. S. Liem}
\IEEEauthorblockA{
c.c.s.liem@tudelft.nl\\
\textit{Delft University of Technology} \\
Delft, The Netherlands \\
}
}

\maketitle

\begin{abstract}
Mutation testing is a well-established technique for assessing a test suite's quality by injecting artificial faults into production code. In recent years, mutation testing has been extended to machine learning (ML) systems, and deep learning (DL) in particular; researchers have proposed approaches, tools, and statistically sound heuristics to determine whether mutants in DL systems are killed or not.
However, as we will argue in this work, questions can be raised to what extent currently used mutation testing techniques in DL are actually in line with the classical interpretation of mutation testing.
We observe that ML model development resembles a test-driven development (TDD) process, in which
a training algorithm (`programmer')
generates a model (program) 
that fits the data points (test data) to labels (implicit assertions), up to a certain threshold. However, considering proposed mutation testing techniques for ML systems under this TDD metaphor, in current approaches, the distinction between production and test code is blurry, and the realism of mutation operators can be challenged. We also consider the fundamental hypotheses underlying classical mutation testing: the competent programmer hypothesis and coupling effect hypothesis. As we will illustrate, these hypotheses do not trivially translate to ML system development, and more conscious and explicit scoping and concept mapping will be needed to truly draw parallels.
Based on our observations, we propose several action points for better alignment of mutation testing techniques for ML with paradigms and vocabularies of classical mutation testing.

\end{abstract}


\section{Introduction}
In many present-day systems,
automated data processing powered by machine learning (ML) techniques has become an essential component.
ML techniques have been important for reaching scale and efficiency in making data-driven predictions
which previously required human judgment. In several situations, ML models---especially those based on Deep Learning (DL) techniques---even have been claimed to perform `better than humans'~\cite{he2015rectifiers,silver2017alphago}. At the same time, this latter claim has met with controversy, as seemingly well-performing ML models were found to make unexpected mistakes that humans would not make~\cite{sturm2014horse, nguyen2015fooled}. 

One could argue that ML procedures effectively are software procedures, and thus, that undesired ML pipeline behavior signals faults or bugs. Therefore, a considerable body of `testing for ML' work has emerged in the software engineering community, building upon established software testing techniques, as extensively surveyed by Ben Braiek \& Khom~\cite{braiek2018testing}, as well as Zhang et al.~\cite{zhangetal2019testing}. While this is a useful development, in this work, we argue that \emph{current testing for ML approaches are not sufficiently explicit about what is being tested exactly}. In particular, in this paper, we focus on mutation testing, a well-established white-box testing technique, which recently has been applied in the context of DL. As we will show, several fundaments of classical mutation testing are currently unclearly or illogically mapped to the ML context. After illustrating this, we will conclude this paper with several concrete action points for improvement.

\section{ML Model Development}
\label{sec:development}
When a (supervised) ML model is being developed, first of all, a domain expert or data scientist will collect a dataset, in which each data point evidences input-output patterns that the ML model should learn to pick up. When input from this dataset is offered to a trained ML model, the model should be capable of yielding the corresponding output.

In contrast to the development of traditional software~\cite{braiek2018testing, zhangetal2019testing}, the process that yields the trained model will partially be generated through automated optimization routines, for which implementations in ML frameworks are usually employed. The data scientist explicitly specifies the type of ML model that should be trained, with any hyperparameters of interest. This choice dictates the architecture of the model to be constructed, and the rules according to which the model will be iteratively fitted to the data during the training phase.

Considering that the training data specifies what output should be predicted for what input, and the goal of model fitting is to match this as well as possible, we argue that \emph{the ML training procedure resembles a test-driven development (TDD) procedure}. Under this perspective, \emph{the training data can be considered as a test suite}, which is defined before system development is initiated. Initially, as training starts, a model $M$ will not yet fit the data well; in other words, too many tests will fail. Therefore, a new, revised version $M'$ of the model will be created. The more tests will be passing, the closer the model will be considered to be to `the correct program'. The process is repeated until convergence, and/or until a user-defined performance threshold will be met.

As soon as model training is finished, the performance of the trained model will be assessed against a test set that was not used during training. As a consequence, performance evaluation of a trained model can be seen as \emph{another TDD procedure}, now \emph{considering the test data as the test suite}. If too many test cases fail, the trained model is unsatisfactory, and will need to be replaced by another, stronger model. Here, the `other, stronger model' can be considered in a black-box fashion; it may be a variant of the previously fitted model, or a completely different model altogether.

Finally, as soon as the satisfactory performance is reached, a trained model will likely be integrated as a component into a larger system. While this integration is outside the scope of the current paper, we would like to note that practical concerns regarding ML faults often are being signaled in this integrated context (e.g.\ unsafe behavior of a self-driving car), but consequently may not only be due to the ML model.

\section{Classical Mutation Testing} 
In `traditional' software testing, mutation testing is an extensively surveyed~\cite{Jia2011survey, papadakis2019mutation, zhu2018systematic}, well-established field, for which early work can be traced back to the 1970s.

\smallskip

\textbf{The process}.
Given a program $P$, mutation testing tools generate program variants (i.e., \textit{mutants}) based on a given set of transformation rules (i.e., \textit{mutation operators}). The mutants correspond to potential (small) mistakes that \textit{competent programmers} could make and that are artificially inserted in the production code. 
If the test suite cannot detect these injected mutants, it may not be effective in discovering real defects.
More precisely, an effective test case should pass when executed against the original program $P$, but fail against a mutant $P'$. In other words, the test should differentiate between the original program and its mutated variant. In this scenario, the mutant $P'$ is said to be \textit{killed}. Instead, the mutant $P
'$ \textit{survives} if the test cases pass on both $P$ and $P'$. The test suite's effectiveness is measured as the ratio between the number of killed mutants and the total number of non-equivalent mutants being generated (\textit{mutation score})~\cite{Jia2011survey}. Therefore, changes (mutants) are applied to the production code, while the test code is left unchanged.
Mutants can only be killed if they relate to tests that passed for the original program.
After mutation analysis, test code can be further improved
by adding/modifying tests and assertions.

\smallskip

\textbf{The hypotheses}.
As reported by Jia and Harman~\cite{Jia2011survey}, mutation testing targets faults in programs that are close to their correct version. 
The connection between (small) mutants and real faults is supported by two fundamental hypotheses: the \textit{Competent Programmer Hypothesis} (CPH)~\cite{offutt1992investigations} and the \textit{Coupling Effect Hypothesis} (CEH)~\cite{demillo1978hints}.
Based on these hypotheses, mutation testing only applies simple syntactic changes/mutants, as they correspond to mistakes that a competent programmer tends to make. Furthermore, each (first-order) mutant $P'$ includes one single syntactic change, since a test case that detects $P'$ will also be able to detect more complex (higher-order) faults coupled to $P'$.
\smallskip

\textbf{Challenges in mutation testing}: Prior studies~\cite{just2014mutants, papadakis2018mutation, andrews2005mutation} showed that mutants are valid substitutes of real faults and, therefore, mutation testing can be used to assess the quality of a test suite. However, mutation testing has some important challenges: the high computation cost and equivalent mutants. 
%
\textit{Equivalent mutants} are mutants that always produce the same output as the original program; therefore, they cannot be killed~\cite{Jia2011survey}. While \textit{program equivalence} is an undecidable problem, researches have proposed various strategies to avoid the generation of equivalent mutants for classical programs~\cite{Jia2011survey}.


\section{Mutation Testing Approaches for ML}
\label{sec:related}
Mutation testing has also been applied to ML software. To demonstrate the effectiveness of metamorphic testing strategies for ML classification algorithm implementations, in Xie et al.~\cite{xie2011testing}, mutation analysis was performed using classical syntactical operators. With the recent rise in popularity of deep learning (DL) techniques, accompanied by concerns about DL robustness, interest in mutation testing has re-emerged as a way to assess the adequacy of test data for deep models. To this end, in 2018, two mutation testing proposals were proposed in parallel: MuNN by Shen et al.~\cite{shen2018munn}, and DeepMutation by Ma et al.~\cite{ma2018deepmutation}, both seeking to propose mutation operators tailored to deep models. Both methods focus on convolutional neural network (CNN) models for `classical' image recognition tasks (MNIST digit recognition in MuNN and DeepMutation, plus CIFAR-10 object recognition in DeepMutation).

MuNN considers an already trained convolutional neural network as production code, and proposes 5 mutation operators on this code, that consider deletions or replacements of neurons, activation functions, bias and weight values.
DeepMutation argues that the `source of faults' in DL-based systems can be caused by faults in data and the training program used for model fitting. As such, the authors propose several \emph{source-level mutation operators}, which generate mutants \emph{before} the actual training procedure is executed. In their turn, the source-level operators are divided into \emph{operators that affect training data} (e.g.\ Label Error, Data Missing)
and \emph{operators that affect model structure}
(e.g.\ Layer Removal).
Besides this, the authors also propose eight \emph{model-level mutation operators}, that operate on weights, neurons, or layers of already-trained models.
These model-level operators can be compared to the operators in MuNN; indeed, one can e.g.\ argue that DeepMutation's Gaussian Fuzzing operator is a way to implement MuNN's Change Weight Value operator, where in both cases, weights in the trained model will be changed.

Noticing that DL training procedures include stochastic components, such that re-training under the same conditions will not yield identical models, and that hyperparameters may affect the effectiveness of mutation operators, Jahangirova \& Tonella revisit the operators of MuNN and DeepMutation from a stochastic perspective~\cite{jahangirova_tonella_2020} and study operator (in)effectiveness, also adding a regression task (steering wheel angle prediction from self-driving car images) to the MNIST and CIFAR-10 classification tasks. Furthermore, the authors study more traditional syntactic mutations on the training program, employing the \texttt{MutPy}
tool. Similarly, Chetouane et al.~\cite{chetouane2019pit} do this for neural network implementations in Java, employing the \texttt{PIT} tool~\cite{pit}. Finally, seeking to expand mutation testing beyond CNN architectures, the DeepMutation++ tool~\cite{hu2019deepmutationpp} both includes DeepMutation's CNN-specific mutation operators and various operators for Recurrent Neural Network architectures.

\section{Was Mutation Testing Truly Applied to ML?}
\label{sec:rethinking}

\begin{lstlisting}[
    language=Java,
    caption={Tests for ML training and performance assessment},
    label={lst:example},
    abovecaptionskip=0pt,
    belowcaptionskip=0pt,
    captionpos=b,
    escapechar=\%,
    float=tp,
    belowskip=-\baselineskip,
    numbers=left,  
    xleftmargin=2em,
    frame=top,
    frame=bottom,
    framexleftmargin=2em
]
%\textcolor{gray}{@Test}%
public void testLearning() {
    List<Image> trainingData = FileUtils.readDataset('training_data');
    Model m = Model.train(trainingData); 
    double counter = 0;
    for (Image image : trainingData){
        String outcome = m.predict(image);
        String expected = image.getGroundTruth();
        if (outcome.equals(expected))
            counter++;
    }
    double accuracy = counter / trainingData.size();
    %\textit{assertTrue}%(accuracy >= 0.9);
}
    
%\textcolor{gray}{@Test}%
public void testPerformance(){
    List<Image> testData = FileUtils.readDataset('test_data');
    Model m = Model.readModel('model_path');
    double counter = 0;
    for (Image image : testData){
        String outcome = m.predict(image);
        String expected = image.getGroundTruth();
        if (outcome.equals(expected))
            counter++;
    }
    double accuracy = counter / testData.size();
    %\textit{assertTrue}%(accuracy >= 0.9);
}
\end{lstlisting}

\subsection{Is the production code being mutated?}
\label{sec:productionortest}
Reconsidering our framing in Section~\ref{sec:development} of ML model development as two TDD procedures (one focusing on model fitting, and one focusing on performance evaluation of a trained model), this leads to two types of production code, with two corresponding (implicit) test suites: respectively, the data referred to as `training data' and `test data' in ML vocabulary.

We can frame the assessments of how well a model has been trained, and how well it performs after training, as separate test cases.  Listing~\ref{lst:example} illustrates this for a fictional image recognition task implemented in Java. The \texttt{testLearning} method assesses the success of a training procedure:  it reads the training set (line 3), runs the training procedure for a given model (line 4), and lets the test pass in case a user-provided success criterion (here: a minimum accuracy threshold) is met (line 14).
%
The \texttt{testPerformance} method runs predictions by an already-trained model on a given dataset, again letting the test pass in case the user-provided success criterion is met. 
Under this framing, we question to what extent the operators proposed for DL testing (Section~\ref{sec:related}) are true mutation operators. In fact, we argue that the category of post-training model-level mutation operators is the only one clearly fitting the mutation testing paradigm, in which production code is altered to assess the quality of the test suite---while still raising questions regarding the fit to the fundamental mutation testing hypotheses, as we will explain in Section~\ref{sec:revisiting}.

Many source-level mutation operators actually target the model's (so the production code's) correctness, rather than the quality of the test suite.
For instance, let us consider the \textit{Label Error} operator, which changes the label for one (or more) data point(s) in the training set. Considering the TDD metaphor for ML training, this operator alters the test suite (training data), rather than the production code that generates or represents the trained model. Furthermore, the changes are applied to assess that the model trained on the original dataset provides statistically better predictions than a model\footnote{Same model, with same training algorithm, and same parameter setting.} trained with
faulty data labels. This therefore appears to be test amplification rather than a mutation testing, and this consideration holds for any proposed data-level operator.

For source-level model structure operators, the situation is ambiguous. The code written by the data scientist to initiate model training may either be seen as production code, or as `the specification that will lead to the true production code' (the trained model). It also should be noted that this `true production code' is yielded by yet another software program implementing the training procedure, that typically is hidden in the ML framework.

\subsection{Revisiting the fundamental mutation testing hypotheses}\label{sec:revisiting}
We argue that the two fundamental hypotheses of mutation testing should be reconsidered for ML: many terms in these hypotheses are not easily mapped to ML contexts.

\vspace*{1mm}
\hspace*{-5mm}
\begin{tikzpicture}
\node [mybox] (box){%
\centering
\begin{minipage}{.465\textwidth}
\textbf{CPH} Competent programmers ``tend to develop programs close to the correct version.''~\cite{offutt1992investigations}
\end{minipage}
};
\end{tikzpicture}%

First of all, \textit{for ML, not all `programmers' are human, and the degree of their competence may not always show in production code.}
As discussed in Sections~\ref{sec:development} and~\ref{sec:related}, when an ML model is to be trained, the general setup for the training procedure will be coded by a human being (often, a data scientist). This program could be considered as human-written production code, but is not the true production code: it triggers an automated optimization procedure (typically previously authored by an ML framework programmer), which will yield the target program (the trained model).

Questions can be raised about \textit{what makes for `realistic' mistakes}. Currently proposed operators on data (which, as we argued in Section~\ref{sec:productionortest}, should be interpreted as operators for test amplification, rather than for mutation testing) largely tackle more mechanical potential processing issues (e.g., inconsistent data traversal and noisy measurement), rather than faults that a human operator would make. Generally, the more interesting faults made during dataset curation can have considerable effects downstream, but are not necessarily encoded in the form of small mistakes. For example, if biased data is fed to the pipeline, this may lead to discriminatory decisions downstream. However, one cannot tell from a single data point whether it resulted from biased curation. Furthermore, the undesired downstream behavior 
will also not be trivially traceable to source code; instead, it rather is a signal of implicit, non-functional requirements not having been met. Thus, higher-level acceptance testing will be needed for problems like this, rather than lower-level mutation testing strategies.

As mentioned in Section~\ref{sec:productionortest}, the post-training model-level operators proposed in literature fit the classical mutation testing paradigm. Yet, some of them are unlikely to represent realistic faults made by the `programmer' (the ML training framework). The way the framework iteratively optimizes the model fit is bound to strict and consistent procedures; thus, it e.g.\ will not happen that subsequent model fitting iterations will suddenly change activation functions or duplicate layers.

Finally, \textit{the concept of `an ML system close to the correct version' is ill-defined.} In contrast to classical software testing setups, where all tests should pass for a program to be considered correct, this target is fuzzier in ML systems. During the training phase, one wants to achieve a good model fit, but explicitly wish to \textit{avoid} a 100\% fit on the training data, since this implies overfitting. As soon as a model has been trained, performance on unseen test data is commonly considered as a metric that can be optimized for. Still, the test data points may not constitute a perfect oracle. As a consequence, perfect performance on the unseen test set may still not guarantee that an ML pipeline behaves as intended from a human, semantic perspective~\cite{bertinettoetal2020, liempanichella2020oracle}. Thus, when considering mutation testing for ML, we actually cannot tell the extent to which the program to mutate is already close to the behavior we ultimately intend.

\vspace*{1mm}
\hspace*{-5mm}
\begin{tikzpicture}
\node [mybox] (box){%
\centering
\begin{minipage}{.465\textwidth}
\textbf{CEH} ``Complex mutants are coupled to simple mutants in such a way that a test dataset that detects all simple mutants in a program will also detect a large percentage of the complex mutant.''~\cite{demillo1978hints}
\end{minipage}
};
\end{tikzpicture}%

\textit{In ML, we still lack a clear sense of what makes for simple or complex mutants}. As noted in~\cite{jahangirova_tonella_2020}, the stochastic nature of training may lead to multiple re-trainings of the same model on the same data leading to models that may differ in weights and output. Also, when considering mutations after a model has been trained, not all neurons and layers have equal roles, so the effect of mutating a single neuron may greatly differ.




\section{Conclusion and future work}
In this paper, we discussed how classical hypotheses and techniques for mutation testing do not always clearly translate to how mutation testing has been applied to DL systems. 
We propose several action points for SE researchers to consider, when seeking to further develop mutation testing for ML systems in general, while staying aligned to the well-established paradigms and vocabularies of classical mutation testing.


\textit{When defining new mutation operators for ML, be explicit about what production and test code are.}  In classical mutation testing, there is a clear distinction between the production code (to be altered by mutation operators) and the test code (to be assessed by mutation testing). This clear distinction must be kept when applying mutation testing to ML systems. 
    
\textit{Clearly settle on what the system under test (SUT) is} when extending software testing techniques to ML systems. Currently, \textit{mutation testing for DL} is a common umbrella for all techniques that target DL models, the training procedure, and the surrounding software artifacts. However, as we argued in Section~\ref{sec:development}, ML model development (both within and outside of DL) involves several development stages with associated artifacts, each requiring dedicated, different testing strategies. Furthermore, where in traditional ML, one test set suffices for a specific ML task (regardless of what ML model is chosen for this task), the mutation testing paradigm implies that the adequacy of a given test suite may differ, depending on the program it is associated with. Thus, depending on the program, a given test suite may need different amplifications.

\textit{Work with experts}
to better determine what `realistic mutants', `realistic faults' and `realistic tests' are. 
Ma et al.~\cite{ma2018deepmutation} rightfully argued that not all faults in DL systems are created by humans, but little insight exists yet on how true, realistic faults can be mimicked. Humbatova et al.~\cite{humbatova2020realfaults} recently proposed a taxonomy of real faults in DL systems, finding that many of these have not been targeted in existing mutation testing approaches yet.
Furthermore, when the mutation score will indicate that a given test set is not sufficiently adequate yet, it also still is an open question how test suites can be improved: test cases need to both meet system demands, while having to be encoded in the form of realistic data points. Here, insights from domain and ML experts will be beneficial.

\textit{Investigate what `a small mutant' is.} This aspect is rather complex in ML: repeating the exact same training process on the exact same data is not mutation, but may yield major model differences, due to stochasticity in training.
Similarly, a small back-propagation step may affect many weights in a DL model. At the same time, given the high dimensionality of data and high amounts of parameters in DL models, other mutations (e.g.\ noise additions to pixels) may be unnoticeable as first-order mutants, and may need to be applied in many places at once to have any effect.

Finally, we wish to point out that \textit{the evolution of ML programs differs from `traditional' software}. In traditional software, revisions are performed iteratively, applying small improvements to previous production code. This is uncommon in ML, where revisions are typically considered at the task level (e.g.\ `find an improved object recognition model'), rather than at the level of production code (e.g.\ `see whether changing neuron connections in a previously-trained deep neural network can improve performance'). Translated to software engineering processes, this is more similar to re-implementing the project many times by different, independent developer teams, than to traditional software development workflows. In such cases, if the production code of trained models will not be expected to evolve, employing black-box testing strategies may overall be more sensible than employing white-box testing strategies. Again, this question requires collaboration and further alignment between software engineering, machine learning, and domain experts, which may open many more interesting research directions.
%
%

\balance
\bibliographystyle{IEEEtran}
\bibliography{references}

\end{document}